\documentclass[12pt]{article}
 \usepackage{epsfig}
 \usepackage{pstricks}
 \usepackage{pst-coil}

\newcommand{\dd}{\mbox{d}}

\newcommand{\mmoo}{{(m_D^{\dg}m_D)_{11}}}

\newcommand{\mmii}{{(m_D^{\dg}m_D)_{ii}}}

\newcommand{\sni}{\widetilde{N_i^c}}
\newcommand{\snone}{\widetilde{N_1^c}}

\def\a{\alpha}
\def\b{\beta}

\def\e{\epsilon}
\def\f{\phi}
\def\g{\gamma}
\def\h{\eta}

\def\j{\psi}

\def\l{\lambda}
\def\m{\mu}
\def\n{\nu}

\def\p{\pi}

\def\r{\rho}
\def\s{\sigma}
\def\t{\tau}

\def\D{\Delta}

\def\G{\Gamma}

\def\ve{\varepsilon}
\def\vf{\varphi}

\def\cl{{\cal L}}

\def\co{{\cal O}}

\def\bo{{\raise.15ex\hbox{\large$\Box$}}}               
\def\pr{\prod}                                          
\def\ltap{\raisebox{-.4ex}{\rlap{$\sim$}} \raisebox{.4ex}{$<$}}   
\def\face{{\raise.2ex\hbox{$\displaystyle \bigodot$}\mskip-2.2mu \llap {$\ddot
        \smile$}}}                                      
\def\dg{\dagger}                                     

\def\wt#1{\widetilde{#1}}                    
\def\VEV#1{\left\langle #1\right\rangle}        
\def\leftrightarrowfill{$\mathsurround=0pt \mathord\leftarrow \mkern-6mu
        \cleaders\hbox{$\mkern-2mu \mathord- \mkern-2mu$}\hfill
        \mkern-6mu \mathord\rightarrow$}       
\def\dvec#1{\vbox{\ialign{##\crcr
        \leftrightarrowfill\crcr\noalign{\kern-1pt\nointerlineskip}
        $\hfil\displaystyle{#1}\hfil$\crcr}}}           

\def\beq{\begin{equation}}
\def\eeq{\end{equation}}

\def\beqx{\begin{displaymath}}
\def\eeqx{\end{displaymath}}

\def\beqa{\begin{eqnarray}}
\def\eeqa{\end{eqnarray}}
\def\NO{\nonumber}

\def\pl#1#2#3{Phys.~Lett.~{\bf B {#1}} (19{#2}) #3}
\def\np#1#2#3{Nucl.~Phys.~{\bf B {#1}} (19{#2}) #3}
\def\prl#1#2#3{Phys.~Rev.~Lett.~{\bf #1} (19{#2}) #3}
\def\pr#1#2#3{Phys.~Rev.~{\bf D {#1}} (19{#2}) #3}

\def\ap#1#2#3{Ann.~of Phys.~{\bf {#1}} (19{#2}) #3}

\def\mpl#1#2#3{Mod.~Phys.~Lett.~{\bf A {#1}} (19{#2}) #3}
\def\nc#1#2#3{Nuovo Cim.~{\bf {#1}} (19{#2}) #3}

\catcode`@=11
\def\@citex[#1]#2{\if@filesw\immediate\write\@auxout{\string\citation{#2}}\fi
  \def\@citea{}\@cite{\@for\@citeb:=#2\do
    {\@citea\def\@citea{,\penalty\@m}\@ifundefined
      {b@\@citeb}{{\bf ?}\@warning
       {Citation `\@citeb' on page \thepage \space undefined}}%
\hbox{\csname b@\@citeb\endcsname}}}{#1}}

\def\citer{\@ifnextchar [{\@tempswatrue\@citexr}{\@tempswafalse\@citexr[]}}
\def\@citexr[#1]#2{\scriptsize 
  \if@filesw\immediate\write\@auxout{\string\citation{#2}}\fi
  \def\@citea{}\@cite{\@for\@citeb:=#2\do
    {\@citea\def\@citea{-\penalty\@m}\@ifundefined
       {b@\@citeb}{{\bf ?}\@warning
       {Citation `\@citeb' on page \thepage \space undefined}}%
\hbox{\csname b@\@citeb\endcsname}}}{#1}\normalsize}
\catcode`@=12

\catcode`\@=11
\long\def\@makefntext#1{ 
\protect\noindent \hbox to 3.2pt {\hskip-.9pt  
$^{{\ninerm\@thefnmark}}$\hfil}#1\hfill} 

\def\thefootnote{\fnsymbol{footnote}}
 \def\@makefnmark{\hbox to 0pt{$^{\@thefnmark}$\hss}}  
        
\def\ps@myheadings{\let\@mkboth\@gobbletwo
\def\@oddhead{\hbox{} 
\rightmark\hfil\ninerm\thepage}   
\def\@oddfoot{}\def\@evenhead{\ninerm\thepage\hfil 
\leftmark\hbox{}}\def\@evenfoot{}
\def\sectionmark##1{}\def\subsectionmark##1{}}

\begin{document}

\newcommand{\symbolfootnote}{\renewcommand{\thefootnote}
        {\fnsymbol{footnote}}}
\renewcommand{\thefootnote}{\fnsymbol{footnote}}
\newcommand{\alphfootnote}
        {\setcounter{footnote}{0}
         \renewcommand{\thefootnote}{\sevenrm\alph{footnote}}}

\newcounter{sectionc}\newcounter{subsectionc}\newcounter{subsubsectionc}
\renewcommand{\section}[1] {\vspace{0.6cm}\addtocounter{sectionc}{1} 
\setcounter{subsectionc}{0}\setcounter{subsubsectionc}{0}\noindent 
        {\bf\thesectionc. #1}\par\vspace{0.4cm}}
\renewcommand{\subsection}[1] {\vspace{0.6cm}\addtocounter{subsectionc}{1} 
        \setcounter{subsubsectionc}{0}\noindent 
        {\it\thesectionc.\thesubsectionc. #1}\par\vspace{0.4cm}}
\renewcommand{\subsubsection}[1] {\vspace{0.6cm}\addtocounter{subsubsectionc}{1}
        \noindent {\rm\thesectionc.\thesubsectionc.\thesubsubsectionc. 
        #1}\par\vspace{0.4cm}}
\newcommand{\nonumsection}[1] {\vspace{0.6cm}\noindent{\bf #1}
        \par\vspace{0.4cm}}
                                                 
\newcounter{appendixc}
\newcounter{subappendixc}[appendixc]
\newcounter{subsubappendixc}[subappendixc]
\renewcommand{\thesubappendixc}{\Alph{appendixc}.\arabic{subappendixc}}
\renewcommand{\thesubsubappendixc}
        {\Alph{appendixc}.\arabic{subappendixc}.\arabic{subsubappendixc}}

\renewcommand{\appendix}[1] {\vspace{0.6cm}
        \refstepcounter{appendixc}
        \setcounter{figure}{0}
        \setcounter{table}{0}
        \setcounter{equation}{0}
        \renewcommand{\thefigure}{\Alph{appendixc}.\arabic{figure}}
        \renewcommand{\thetable}{\Alph{appendixc}.\arabic{table}}
        \renewcommand{\theappendixc}{\Alph{appendixc}}
        \renewcommand{\theequation}{\Alph{appendixc}.\arabic{equation}}
        \noindent{\bf Appendix \theappendixc #1}\par\vspace{0.4cm}}
\newcommand{\subappendix}[1] {\vspace{0.6cm}
        \refstepcounter{subappendixc}
        \noindent{\bf Appendix \thesubappendixc. #1}\par\vspace{0.4cm}}
\newcommand{\subsubappendix}[1] {\vspace{0.6cm}
        \refstepcounter{subsubappendixc}
        \noindent{\it Appendix \thesubsubappendixc. #1}
        \par\vspace{0.4cm}}

\newcommand{\bibit}{\it}
\newcommand{\bibbf}{\bf}
\renewenvironment{thebibliography}[1]
        {\begin{list}{\arabic{enumi}.}
        {\usecounter{enumi}\setlength{\parsep}{0pt} 
\setlength{\leftmargin 1.25cm}{\rightmargin 0pt} 
         \setlength{\itemsep}{0pt} \settowidth
        {\labelwidth}{#1.}\sloppy}}{\end{list}}

\topsep=0in\parsep=0in\itemsep=0in
\parindent=1.5pc

\newcounter{itemlistc}
\newcounter{romanlistc}
\newcounter{alphlistc}
\newcounter{arabiclistc}
\newenvironment{itemlist}
        {\setcounter{itemlistc}{0}
         \begin{list}{$\bullet$}
        {\usecounter{itemlistc}
         \setlength{\parsep}{0pt}
         \setlength{\itemsep}{0pt}}}{\end{list}}

\newenvironment{romanlist}
        {\setcounter{romanlistc}{0}
         \begin{list}{$($\roman{romanlistc}$)$}
        {\usecounter{romanlistc}
         \setlength{\parsep}{0pt}
         \setlength{\itemsep}{0pt}}}{\end{list}}

\newenvironment{alphlist}
        {\setcounter{alphlistc}{0}
         \begin{list}{$($\alph{alphlistc}$)$}
        {\usecounter{alphlistc}
         \setlength{\parsep}{0pt}
         \setlength{\itemsep}{0pt}}}{\end{list}}

\newenvironment{arabiclist}
        {\setcounter{arabiclistc}{0}
         \begin{list}{\arabic{arabiclistc}}
        {\usecounter{arabiclistc}
         \setlength{\parsep}{0pt}
         \setlength{\itemsep}{0pt}}}{\end{list}}

\newcommand{\fcaption}[1]{
        \refstepcounter{figure}
        \setbox\@tempboxa = \hbox{\tenrm Fig.~\thefigure. #1}
        \ifdim \wd\@tempboxa > 6in
           {\begin{center}
        \parbox{6in}{\tenrm\baselineskip=12pt Fig.~\thefigure. #1 }
            \end{center}}
        \else
             {\begin{center}
             {\tenrm Fig.~\thefigure. #1}
              \end{center}}
        \fi}

\newcommand{\tcaption}[1]{
        \refstepcounter{table}
        \setbox\@tempboxa = \hbox{\tenrm Table~\thetable. #1}
        \ifdim \wd\@tempboxa > 6in
           {\begin{center}
        \parbox{6in}{\tenrm\baselineskip=12pt Table~\thetable. #1 }
            \end{center}}
        \else
             {\begin{center}
             {\tenrm Table~\thetable. #1}
              \end{center}}
        \fi}

\def\@citex[#1]#2{\if@filesw\immediate\write\@auxout
        {\string\citation{#2}}\fi
\def\@citea{}\@cite{\@for\@citeb:=#2\do
        {\@citea\def\@citea{,}\@ifundefined
        {b@\@citeb}{{\bf ?}\@warning
        {Citation `\@citeb' on page \thepage \space undefined}}
        {\csname b@\@citeb\endcsname}}}{#1}}

\newif\if@cghi
\def\cite{\@cghitrue\@ifnextchar [{\@tempswatrue
        \@citex}{\@tempswafalse\@citex[]}}
\def\citelow{\@cghifalse\@ifnextchar [{\@tempswatrue
        \@citex}{\@tempswafalse\@citex[]}}
\def\@cite#1#2{{$\null^{#1}$\if@tempswa\typeout
        {IJCGA warning: optional citation argument 
        ignored: `#2'} \fi}}
\newcommand{\citeup}{\cite}

\def\fnm#1{$^{\mbox{\scriptsize #1}}$}
\def\fnt#1#2{\footnotetext{\kern-.3em
        {$^{\mbox{\sevenrm #1}}$}{#2}}}

\font\twelvebf=cmbx10 scaled\magstep 1
\font\twelverm=cmr10 scaled\magstep 1
\font\twelveit=cmti10 scaled\magstep 1
\font\elevenbfit=cmbxti10 scaled\magstephalf
\font\elevenbf=cmbx10 scaled\magstephalf
\font\elevenrm=cmr10 scaled\magstephalf
\font\elevenit=cmti10 scaled\magstephalf
\font\bfit=cmbxti10
\font\tenbf=cmbx10
\font\tenrm=cmr10
\font\tenit=cmti10
\font\ninebf=cmbx9
\font\ninerm=cmr9
\font\nineit=cmti9
\font\eightbf=cmbx8
\font\eightrm=cmr8
\font\eightit=cmti8

\date{}
\title{
{\large\rm DESY 97-189}\hfill{\large\tt ISSN 0418-9833}\\
{\large\rm November 1997}\hfill{\large\tt hep-ph/9711208}\vspace*{3cm}\\
{\bf Baryon Asymmetry of the Universe\\ and Lepton Mixing}
\thanks{presented at the 4th. Colloque Cosmologie, Paris, June 1997}}
\author{W. Buchm\"uller and M. Pl\"umacher\\
\vspace{3.0\baselineskip}                                               
{\normalsize\it Deutsches Elektronen-Synchrotron DESY, 22603 Hamburg, Germany}
\vspace*{1cm}\\                     
}        

\maketitle

\thispagestyle{empty}

\begin{abstract}
\noindent
Baryogenesis appears to require lepton number violation. This is
naturally realized in extensions of the standard model containing
right-handed neutrinos. We discuss the generation of a baryon
asymmetry by the out-of-equilibrium decay of heavy Majorana neutrinos
in these models, without and with supersymmetry. All relevant lepton
number violating scattering processes which can inhibit the generation
of an asymmetry are taken into account. We assume a similar pattern of
mixings and masses for neutrinos and up-type quarks, as suggested by
SO(10) unification. This implies that $B-L$ is broken at the
unification scale $\Lambda_{\mbox{\scriptsize GUT}}\sim 10^{16}\;$GeV,
if $m_{\n_\m} \sim 3\cdot10^{-3}\;$eV, as preferred by the MSW
solution to the solar neutrino deficit. The observed baryon asymmetry
is then obtained without any fine tuning of parameters.

\end{abstract}

\newpage

\section{Sphaleron transitions and thermal equilibrium in the early universe}

\begin{figure}[t]
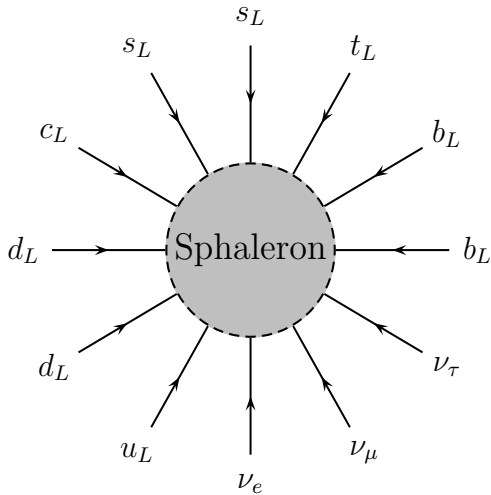

    \begin{center}
      \scaleboxto(7,7){
        \parbox[c]{9cm}{ \begin{center}
     \pspicture*(-0.50,-2.5)(8.5,6.5)
     \psset{linecolor=lightgray}
     \qdisk(4,2){1.5cm}
     \psset{linecolor=black}
     \pscircle[linewidth=1pt,linestyle=dashed](4,2){1.5cm}
     \rput[cc]{0}(4,2){\scalebox{1.5}{Sphaleron}}
     \psline[linewidth=1pt](5.50,2.00)(7.50,2.00)
     \psline[linewidth=1pt](5.30,2.75)(7.03,3.75)
     \psline[linewidth=1pt](4.75,3.30)(5.75,5.03)
     \psline[linewidth=1pt](4.00,3.50)(4.00,5.50)
     \psline[linewidth=1pt](3.25,3.30)(2.25,5.03)
     \psline[linewidth=1pt](2.70,2.75)(0.97,3.75)
     \psline[linewidth=1pt](2.50,2.00)(0.50,2.00)
     \psline[linewidth=1pt](2.70,1.25)(0.97,0.25)
     \psline[linewidth=1pt](3.25,0.70)(2.25,-1.03)
     \psline[linewidth=1pt](4.00,0.50)(4.00,-1.50)
     \psline[linewidth=1pt](4.75,0.70)(5.75,-1.03)
     \psline[linewidth=1pt](5.30,1.25)(7.03,0.25)
     \psline[linewidth=1pt]{<-}(6.50,2.00)(6.60,2.00)
     \psline[linewidth=1pt]{<-}(6.17,3.25)(6.25,3.30)
     \psline[linewidth=1pt]{<-}(5.25,4.17)(5.30,4.25)
     \psline[linewidth=1pt]{<-}(4.00,4.50)(4.00,4.60)
     \psline[linewidth=1pt]{<-}(2.75,4.17)(2.70,4.25)
     \psline[linewidth=1pt]{<-}(1.83,3.25)(1.75,3.30)
     \psline[linewidth=1pt]{<-}(1.50,2.00)(1.40,2.00)
     \psline[linewidth=1pt]{<-}(1.83,0.75)(1.75,0.70)
     \psline[linewidth=1pt]{<-}(2.75,-0.17)(2.70,-0.25)
     \psline[linewidth=1pt]{<-}(4.00,-0.50)(4.00,-0.60)
     \psline[linewidth=1pt]{<-}(5.25,-0.17)(5.30,-0.25)
     \psline[linewidth=1pt]{<-}(6.17,0.75)(6.25,0.70)
     \rput[cc]{0}(8.00,2.00){\scalebox{1.3}{$b_L$}}
     \rput[cc]{0}(7.46,4.00){\scalebox{1.3}{$b_L$}}
     \rput[cc]{0}(6.00,5.46){\scalebox{1.3}{$t_L$}}
     \rput[cc]{0}(4.00,6.00){\scalebox{1.3}{$s_L$}}
     \rput[cc]{0}(2.00,5.46){\scalebox{1.3}{$s_L$}}
     \rput[cc]{0}(0.54,4.00){\scalebox{1.3}{$c_L$}}
     \rput[cc]{0}(0.00,2.00){\scalebox{1.3}{$d_L$}}
     \rput[cc]{0}(0.54,0.00){\scalebox{1.3}{$d_L$}}
     \rput[cc]{0}(2.00,-1.46){\scalebox{1.3}{$u_L$}}
     \rput[cc]{0}(4.00,-2.00){\scalebox{1.3}{$\nu_e$}}
     \rput[cc]{0}(6.00,-1.46){\scalebox{1.3}{$\nu_{\mu}$}}
     \rput[cc]{0}(7.46,0.00){\scalebox{1.3}{$\nu_{\tau}$}}
     \endpspicture
\end{center}}
      }
    \end{center}
    \caption{\it One of the 12-fermion processes which are in thermal 
      equilibrium in the high-temperature phase of the standard model.
      \label{fig_sphal}
    }
  \end{figure}

  Due to the chiral nature of the weak interactions baryon number
  ($B$) and lepton number ($L$) are not conserved in the standard
  model (SM)\cite{thoo}. At zero temperature this has no observable
  effect due to the smallness of the weak coupling. However, as the
  temperature approaches the critical temperature $T_c$ of the
  electroweak phase transition, $B$ and $L$ violating processes come
  into thermal equilibrium\cite{krs}. Their rate is determined by the
  free energy of sphaleron-type field configurations which carry
  topological charge. In the standard model they induce an effective
  interaction of all left-handed fermions (cf. fig.~\ref{fig_sphal})
  which violates baryon and lepton number by three units, 
  \beq 
    \D B = \D L = 3\;. \label{sphal1}
  \eeq
  
  Sphaleron processes have a profound effect on the generation of the
  cosmological baryon asymmetry.  Eq.~\ref{sphal1} suggests that any
  $B+L$ asymmetry generated before the electroweak phase transition,
  i.e., at temperatures $T>T_c$, will be washed out. However, since
  only left-handed fields couple to sphalerons, a non-zero value of
  $B+L$ can persist in the high-temperature, symmetric phase if there
  exists a non-vanishing $B-L$ asymmetry.
  
  This is most easily seen in an analysis of the chemical potentials
  involved in this problem\cite{chem}.  The gauge bosons of unbroken
  gauge symmetries have vanishing chemical potentials. In the SM with
  $N_F$ fermion generations and $N_H$ Higgs doublets $\f_i$ we have
  $2N_F$ left-handed quark and lepton doublets $q_{iL}=(u_{iL},
  d_{iL})$ and $l_{iL}=(\n_{iL},e_{iL})$, $3N_F$ right-handed quark
  and charged lepton singlets $u_{iR}$, $d_{iR}$ and $e_{iR}$ and
  $2N_H$ neutral and charged Higgs fields $\vf^0_i$ and $\vf^-_i$.
  However, not all of the corresponding chemical potentials are
  independent. Cabibbo mixing between the quarks will balance out the
  chemical potentials of the various up- and down-quark states,
  respectively. If in addition the mixing between the Higgs doublets
  is strong enough all the Higgs fields $\vf^0_i$ and $\vf^-_i$ have
  the same chemical potentials $\m_0$ and $\m_-$.  In thermal
  equilibrium, perturbative electroweak interactions yield the
  relations,
  \beqa
    W^-\leftrightarrow \vf^-+\vf^0\;&:&\qquad \m_0=-\m_-\;,
      \label{first_condition}\\
    W^-\leftrightarrow\overline{u_L}+d_L\;&:&\qquad
      \m_{d_L}=\m_{u_L}\;,\NO\\
    W^-\leftrightarrow\overline{\n_{iL}}+e_{iL}\;&:&\qquad
      \m_{ie_L}=\m_{i\n}\;,\NO\\
    \vf^0\leftrightarrow \overline{u_L}+u_R\;&:&\qquad
      \m_{u_R}=\m_{u_L}+\m_0\;,\NO\\
    \vf^0\leftrightarrow \overline{d_R}+d_L\;&:&\qquad
      \m_{d_R}=\m_{u_L}-\m_0\;,\NO\\
    \vf^0\leftrightarrow \overline{e_{iR}}+e_{iL}\;&:&\qquad
      \m_{ie_R}=\m_{i\n}-\m_0\;.\NO
  \eeqa 
  The sphaleron processes (cf.~fig.~\ref{fig_sphal}) yield the 
  additional condition
  \beq
    N_F(\m_{u_L}+2\m_{d_L})+\sum\limits_{i=1}^{N_F}\m_{i\n}=0
     \;.\label{last_condition}
  \eeq
  In thermal equilibrium, if all the chemical potentials are small
  compared to the temperature, one obtains for the baryon number $B$ and 
  the lepton number $L$,
  \beqa
    \langle B\rangle_T &=& {n_B-n_{\overline{B}}\over s}=
      {15\over\p^2g_*}{1\over T}\,
      N_F\,(\m_{u_L}+\m_{u_R}+\m_{d_L}+\m_{d_R})\;,\\[1ex]
    \langle L \rangle_T &=& {n_L-n_{\overline{L}}\over s}=
      {15\over\p^2g_*}{1\over T}\,
      \,\sum\limits_{i=1}^{N_F}(\m_{ie_L}+\m_{ie_R}+\m_{i\n})\;,
  \eeqa
  where $g_*$ is the number of relativistic degrees of freedom and $s$
  is the entropy density of the universe. {}From eqs. 
  (\ref{first_condition})-(\ref{last_condition}) it follows that $B$ is
  proportional to $B-L$,
  \beq
    \langle B \rangle_T =C\; \langle B-L \rangle_T = {C\over C-1}\; 
    \langle L \rangle_T
    \quad,\quad C={8N_F+4N_H\over 22N_F+13N_H}\;.
  \eeq
  In the SM, with $N_F=3$ and $N_H=1$ one has $C={28\over79}$. 
  
  We conclude that $B-L$ violation is needed to obtain a non-vanishing
  baryon asymmetry. In the standard model, as well as its
  supersymmetric version and its unified extensions based on the gauge
  group SU(5), $B-L$ is a conserved quantity. Hence, no baryon
  asymmetry can be generated dynamically in these models.
  
  In principle, this conclusion could be avoided if the baryon
  asymmetry were produced directly in a first-order electroweak phase
  transition\cite{phase}. However, a detailed study of the
  thermodynamics indicates that the phase transition in the SM is too
  weak for baryogenesis\cite{jansen}. In the minimal supersymmetric
  extension of the standard model (MSSM) such a scenario is still
  conceivable for a limited range of parameters\cite{susy_ewpt}.

\clearpage
\section{Standard model with right-handed neutrinos}

  The cosmological baryon asymmetry appears to require $B-L$
  violation, and therefore $L$ violation. Lepton number violation is
  naturally realized by adding right-handed Majorana neutrinos to the
  standard model.  Heavy right-handed Majorana neutrinos, whose
  existence is predicted by theories based on gauge groups containing
  SO(10)\cite{so10}, can also explain the smallness of the light
  neutrino masses via the see-saw mechanism\cite{seesaw}.

  The most general lagrangian for couplings and masses of charged
  leptons and neutrinos is given by 
  \beq\label{yuk}
    \cl_Y = -\overline{l_L}\,\wt{\f}\,g_l\,e_R
            -\overline{l_L}\,\f\,g_{\n}\,\n_R
            -{1\over2}\,\overline{\n^C_R}\,M\,\n_R
            +\mbox{ h.c.}\;.
  \eeq
  The vacuum expectation value of the Higgs field $\VEV{\vf^0}=v\ne0$
  generates Dirac masses $m_l$ and $m_D$ for charged leptons and
  neutrinos,
  \beq
     m_l=g_l\,v \quad, \qquad m_D=g_{\n}\,v\;,
  \eeq  
  which are assumed to be much smaller than the Majorana masses $M$.
  This yields light and heavy neutrino mass eigenstates
  \beq
     \n\simeq K^{\dg}\n_L+\n_L^C K\quad,\qquad
     N\simeq\n_R+\n_R^C\, ,
  \eeq
  with masses
  \beq
     m_{\n}\simeq- K^{\dg}m_D{1\over M}m_D^T K^*\,
     \quad,\quad  m_N\simeq M\, .
     \label{seesaw}
  \eeq
  Here $K$ is a unitary matrix which relates weak and mass eigenstates. 

  \begin{figure}[b]
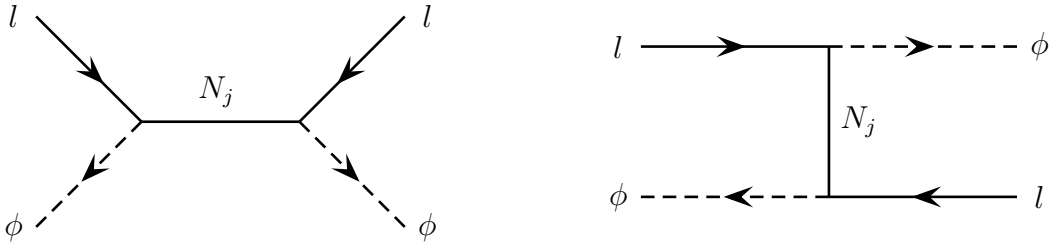

     \begin{center}
  \pspicture[0.5](1,0.3)(8,3.5)
    \psline[linewidth=1pt,linestyle=dashed](1.6,0.6)(3,2)
    \psline[linewidth=2pt]{<-}(2.2,1.2)(2.3,1.3)
    \rput[cc]{0}(1.3,0.6){$\displaystyle \f$}
    \psline[linewidth=1pt](1.6,3.4)(3,2)
    \psline[linewidth=2pt]{->}(2.4,2.6)(2.5,2.5)
    \rput[cc]{0}(1.3,3.4){$\displaystyle l$}
    \psline[linewidth=1pt](3,2)(5.1,2)
    \rput[cc]{0}(4,2.4){$\displaystyle N_j$}
    \psline[linewidth=1pt](5.1,2)(6.5,3.4)
    \psline[linewidth=2pt]{->}(5.7,2.6)(5.6,2.5)
    \rput[cc]{0}(6.8,3.4){$\displaystyle l$}
    \psline[linewidth=1pt,linestyle=dashed](5.1,2)(6.5,0.6)
    \psline[linewidth=2pt]{->}(5.8,1.3)(5.9,1.2)
    \rput[cc]{0}(6.8,0.6){$\displaystyle \f$}
    \endpspicture
    \hspace{1cm}
    \pspicture[0.5](8.5,0.3)(15,3.5)
    \psline[linewidth=1pt](9,3)(11.5,3)
    \psline[linewidth=2pt]{->}(10.3,3)(10.4,3)
    \rput[cc]{0}(8.7,3){$\displaystyle l$}
    \psline[linewidth=1pt,linestyle=dashed](11.5,3)(14,3)
    \psline[linewidth=2pt]{->}(12.8,3)(12.9,3)
    \rput[cc]{0}(14.3,3){$\displaystyle \f$}
    \psline[linewidth=1pt](11.5,3)(11.5,1)
    \rput[cc]{0}(11.9,2){$\displaystyle N_j$}
    \psline[linewidth=1pt,linestyle=dashed](9,1)(11.5,1)
    \psline[linewidth=2pt]{<-}(10.1,1)(10.2,1)
    \rput[cc]{0}(8.7,1){$\displaystyle \f$}
    \psline[linewidth=1pt](11.5,1)(14,1)
    \psline[linewidth=2pt]{<-}(12.6,1)(12.7,1)
    \rput[cc]{0}(14.3,1){$\displaystyle l$}
  \endpspicture
\end{center}
     \caption{\it Lepton number violating lepton Higgs scattering
       \label{lept_fig}}
  \end{figure}

  The exchange of heavy Majorana neutrinos generates an effective
  $\Delta L=2$ interaction at low energies (cf.~fig.~\ref{lept_fig}),
  \beq
  \cl_{\Delta L=2} = {1\over 2}\overline{l_L}\,\f\,g_{\n}\,{1\over M}\,
         g_{\n}^T\,\f\,l_L^c +\mbox{ h.c.}\;.\label{intl2}
  \eeq
  At finite temperature the corresponding $\Delta L=2$ processes take
  place with the rate\cite{etc2}
  \beq
    \Gamma_{\Delta L=2} (T) = {1\over \pi^3}\,{T^3\over v^4}\, 
    \sum\limits_i m_{\n i}^2\, .
  \eeq
  In thermal equilibrium the interaction (\ref{intl2}) implies
  \beq
    \m_0 = \m_{ie_L} = \m_{i\n}\; ,
  \eeq
  and therefore
  \beq
    \langle B \rangle_T = \langle B - L \rangle_T = 0 \; .
  \eeq
  To avoid this conclusion, the $\Delta L=2$ interaction (\ref{intl2})
  must not reach thermal equilibrium, which imposes an upper bound on
  the light neutrino masses $m_\n$.

  \begin{figure}[t]
    \input{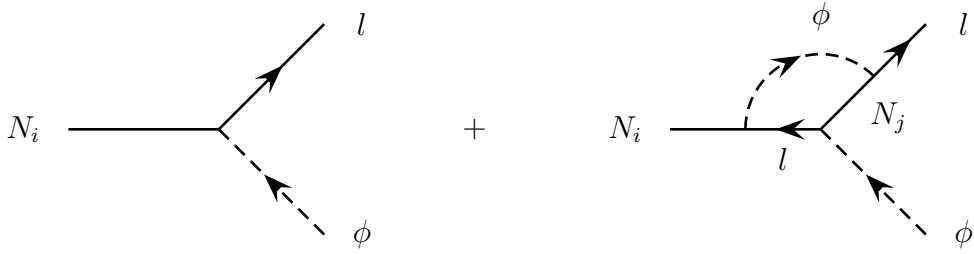}
  \caption{Contributions to the decay of a heavy Majorana neutrino
           \label{decay_fig} }
  \end{figure}
  
  The right-handed neutrinos, whose exchange may erase any lepton
  asymmetry, can also generate a lepton asymmetry by means of
  out-of-equilibrium decays\cite{fy}. This lepton asymmetry is then
  partially transformed into a baryon asymmetry by the sphaleron
  processes.  The decay width of $N_i$ in its rest frame reads at tree
  level (cf.~fig.~\ref{decay_fig}),
  \beqa
    \G_{Di}&=&\G_{rs}\left(N^i\to\f^c+l\right)+
    \G_{rs}\left(N^i\to\f+l^c\right)\NO\\
           &=&{1\over8\p}{(m_D^{\dg}m_D)_{ii}\over v^2}M_i\;.
    \label{decay}
  \eeqa
  The decay width is closely related to the light neutrino masses
  which are therefore constrained by the out-of-equilibrium condition.
  Requiring $\Gamma_{Di}< H|_{T=Mi}$, where $H$ is the Hubble
  parameter, a rough estimate yields\cite{fisch}
  \beq
    m_{\n i} < \ 10^{-3}\, \mbox{eV}\;.
  \eeq
  The detailed calculations described later will be consistent with
  this estimate.

  Interference between the tree-level amplitude and the one-loop vertex
  correction (cf.~fig.~\ref{decay_fig}) yields the $CP$ asymmetry
  \beqa
    \ve_i&=&{\Gamma(N_i\rightarrow l \f^c)-\Gamma(N_i\rightarrow l^c \f)\over
             \Gamma(N_i\rightarrow l \f^c)+\Gamma(N_i\rightarrow l^c \f)}\NO\\
         &=&{1\over8\pi v^2}\;{1\over\left(m_D^{\dag}m_D\right)_{ii}}
      \sum\limits_j
      \mbox{Im}\left[\left(m_D^{\dag}m_D\right)_{ij}^2\right]\,f
      \left({M_j^2\over M_i^2}\right)\label{cpasymm}\;,
  \eeqa
  where
  \beq
    f(x)=\sqrt{x}\left[1-(1+x)\ln\left({1+x\over x}\right)\right]\;.
  \eeq
  Note, that self-energy corrections do not contribute to CP 
  asymmetries\cite{bp_new}.

  {}From the CP asymmetry (\ref{cpasymm}) one obtains a rough estimate of
  the baryon asymmetry (cf.~\cite{kw})
  \beq
    Y_{B-L} = {n_{B-L}\over s} \sim {\ve\over g_*}\;,
    \label{esti}
  \eeq
  where the effective number of degrees of freedom $g_* \simeq 100$ in
  the SM. This estimate is useful as an upper bound on the generated
  baryon asymmetry. However, it is easily too large by a factor
  $\co(100)$.  In order to obtain a more accurate result, one has to
  solve the Boltzmann equations.
  
\section{Boltzmann equations and scattering processes}

  In a quantitative analysis of baryogenesis one has to integrate the
  relevant set of Boltzmann equations which are treated in some
  approximation\cite{kw}.  One usually neglects the difference between
  Bose and Fermi statistics so that the equilibrium phase space
  density of a particle $\j$ with mass $m_{\j}$ is given by
  Maxwell-Boltzmann statistics,
  \beq
     f^{eq}_{\j}\left(E_{\j},T\right)=\mbox{e}^{-E_{\j}/T}\;.
  \eeq
  The corresponding particle density is
  \beq
     n_{\j}(T)={g_{\j}\over(2\p)^3}\int\dd^3p_{\j}\,f_{\j}\;,
  \eeq
  where $g_{\j}$ is the number of internal degrees of freedom. The
  number of particles $Y_{\j}$ in a comoving volume element is given
  by the ratio of $n_{\j}$ and the entropy density $s$. If the
  universe expands isentropically, $Y_{\j}$ is not affected by the
  expansion, i.e.\ $Y_{\j}$ can only be changed by interactions. 
  
  One distinguishes elastic and inelastic scatterings.  Elastic
  scatterings only affect the phase space densities of the particles
  but not the number densities, whereas inelastic scatterings do change
  the number densities. If elastic scatterings do occur at a higher
  rate than inelastic scatterings one can assume kinetic equilibrium,
  i.e., the phase space density is
  \beq
     f_{\j}(E_{\j},T)={n_{\j}\over n_{\j}^{eq}}\mbox{e}^{-E_{\j}/T}\;.
  \eeq
  The Boltzmann equation, which describes the evolution of $Y_{\j}$ with 
  temperature, then reads
  \beqa
    {\mbox{d}Y_{\j}\over\mbox{d}z}&=&-{z\over sH\left(m_{\j}\right)}
    \sum\limits_{a,i,j,\ldots}\left[{Y_{\j}Y_a\ldots\over
     Y_{\j}^{eq}Y_a^{eq}\ldots}\,\g^{eq}\left(\j+a+\ldots\to 
     i+j+\ldots\right)-\right.\NO\\[1ex]
     &&\qquad\qquad\qquad\left.-{Y_iY_j\ldots\over 
     Y_i^{eq}Y_j^{eq}\ldots}
     \,\g^{eq}\left(i+j+\ldots\to\j+a+\ldots\right)\right]\;.
    \label{7}
  \eeqa
  Here $z=m_{\j}/T$ and $H\left(m_{\j}\right)$ is the Hubble
  parameter at $T=m_{\j}$. 

  \begin{figure}[b]
     \input{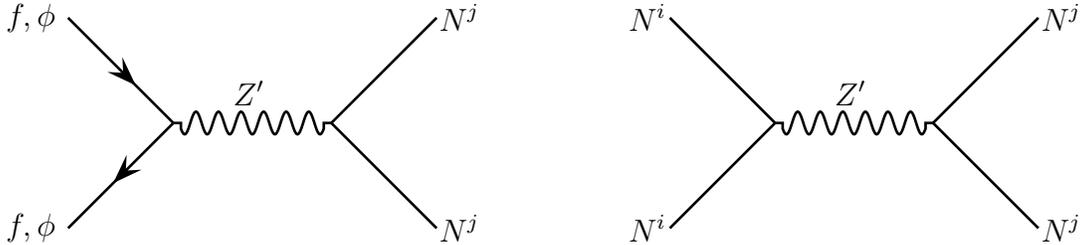}
     \caption{\it Lepton number conserving processes mediated by the
       new neutral gauge boson $Z'$.
      \label{Lcons_fig}}
  \end{figure}
  
  The right-hand side of eq.~(\ref{7}) describes the interactions in
  which a $\j$ particle takes part, where $\g^{eq}$ is the space time
  density of scatterings in thermal equilibrium.  In a dilute gas we
  only have to take into account decays, two-particle scatterings and
  the corresponding back reactions. For a decay one has\cite{luty}
  \beq
     \g_D:=\g^{eq}(\j\to i+j+\ldots)=
     n^{eq}_{\j}{\mbox{K}_1(z)\over\mbox{K}_2(z)}\,\tilde{\G}_{rs}\;,
  \eeq
  where K$_1$ and K$_2$ are modified Bessel functions and
  $\tilde{\G}_{rs}$ is the usual decay width in the rest system of
  the decaying particle. The ratio of the Bessel functions is a
  time dilatation factor. 

  \begin{figure}[t]
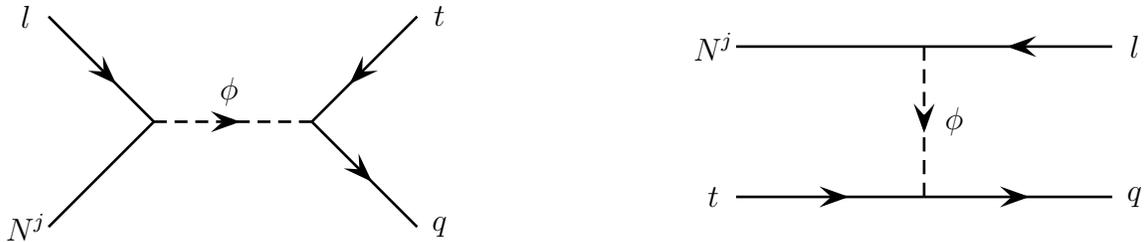

      \begin{center}
  \pspicture[0.5](1.0,0.3)(7.3,3.7)
    \psline[linewidth=1pt](1.6,0.6)(3,2)
    \rput[cc]{0}(1.3,0.6){$\displaystyle N^j$}
    \psline[linewidth=1pt](1.6,3.4)(3,2)
    \psline[linewidth=2pt]{->}(2.4,2.6)(2.5,2.5)
    \rput[cc]{0}(1.3,3.4){$\displaystyle l$}
    \psline[linewidth=1pt,linestyle=dashed](3,2)(5.1,2)
    \psline[linewidth=2pt]{->}(4.05,2)(4.15,2)
    \rput[cc]{0}(4,2.4){$\displaystyle \f$}
    \psline[linewidth=1pt](5.1,2)(6.5,3.4)
    \psline[linewidth=2pt]{->}(5.7,2.6)(5.6,2.5)
    \rput[cc]{0}(6.8,3.4){$\displaystyle t$}
    \psline[linewidth=1pt](5.1,2)(6.5,0.6)
    \psline[linewidth=2pt]{->}(5.8,1.3)(5.9,1.2)
    \rput[cc]{0}(6.8,0.6){$\displaystyle q$}
    \endpspicture
    \hspace{\fill}
    \pspicture[0.5](8.5,0.3)(14.5,3.7)
    \psline[linewidth=1pt](9,3)(11.5,3)
    \rput[cc]{0}(8.7,3){$\displaystyle N^j$}
    \psline[linewidth=1pt](11.5,3)(14,3)
    \psline[linewidth=2pt]{<-}(12.6,3)(12.7,3)
    \rput[cc]{0}(14.3,3){$\displaystyle l$}
    \psline[linewidth=1pt,linestyle=dashed](11.5,3)(11.5,1)
    \psline[linewidth=2pt]{->}(11.5,1.95)(11.5,1.85)
    \rput[cc]{0}(11.9,2){$\displaystyle \f$}
    \psline[linewidth=1pt](9,1)(11.5,1)
    \psline[linewidth=2pt]{->}(10.3,1)(10.5,1)
    \rput[cc]{0}(8.7,1){$\displaystyle t$}
    \psline[linewidth=1pt](11.5,1)(14,1)
    \psline[linewidth=2pt]{->}(12.8,1)(12.9,1)
    \rput[cc]{0}(14.3,1){$\displaystyle q$}
  \endpspicture
\end{center}
      \caption{\it Lepton number violating neutrino top-quark scattering
       \label{top_fig}} 
  \end{figure}

  If one neglects $CP$ violating effects the same reaction density
  describes the inverse decays,
  \beq
     \g_{ID}:=\g^{eq}(i+j+\ldots\to\j)=\g_D\;.
  \eeq
  For two body scattering one has
  \beq
     \g^{eq}({\j}+a\leftrightarrow i+j+\ldots)=
     {T\over64\p^4}\int\limits_{\left(m_{\j}+m_a\right)^2}^{\infty}
     \hspace{-0.5cm}\dd s\,\hat{\s}(s)\,\sqrt{s}\,
     \mbox{K}_1\left({\sqrt{s}\over T}\right)\;,
  \eeq 
  where $s$ is the squared centre of mass energy, and the reduced
  cross section $\hat{\s}(s)$ for the process ${\j}+a\to i+j+\ldots$
  is related to the usual total cross section $\s(s)$ by
  \beq
    \hat{\s}(s)={8\over s}
    \left[\left(p_{\j}\cdot p_a\right)^2-m_{\j}^2m_a^2\right]\,\s(s)\;.
  \eeq
  In kinetic equilibrium, contributions from elastic scatterings drop
  out of eq.~(\ref{7}). 
  
  Consider now the various processes which are relevant in the
  leptogenesis scenario. In order to obtain a lepton asymmetry of the
  correct order of magnitude, the right-handed neutrinos have to be
  numerous before decaying, i.e., they have to be in thermal
  equilibrium at high temperatures. The Yukawa interactions
  (\ref{yuk}) are too weak to achieve this and additional interactions
  are therefore needed.  Since right-handed neutrinos are a necessary
  ingredient of SO(10) unified theories, it is natural to consider
  leptogenesis within an extended gauge model, contained in a SO(10)
  GUT. The minimal extension of the standard model is based on the
  gauge group
  \beq
    G=\mbox{SU}(3)_C\times\mbox{SU}(2)_L\times\mbox{U}(1)_Y\times
     \mbox{U}(1)_{Y'}\ \subset\ \mbox{SO}(10)\;.
  \eeq
  Here $U(1)_{Y'}$, and therefore $B-L$, is spontaneously broken, and
  the breaking scale is related to the heavy neutrino masses.  The
  additional neutral gauge boson $Z'$ accounts for pair creation and
  annihilation processes and for flavour transitions between heavy
  neutrinos of different generations (cf.~fig.~\ref{Lcons_fig}). For
  appropriately chosen parameters these processes generate an
  equilibrium distribution of heavy neutrinos at high
  temperatures\cite{pluemi}.
  
  Of crucial importance are the $\Delta L=2$ lepton number violating
  scatterings shown in fig.~\ref{lept_fig} which, if too strong, erase
  any lepton asymmetry. Similarly, the $\Delta L=1$ lepton number
  violating neutrino top-quark scatterings shown in fig.~\ref{top_fig}
  have to be taken into account because of the large top Yukawa
  coupling. Finally, the heavy neutrino decays
  (cf.~fig.~\ref{decay_fig}) as well as the inverse decays have to be
  incorporated in the Boltzmann equations.
  
  Based on these equations the resulting lepton and baryon asymmetries
  can be evaluated, and it is known that the observed cosmological
  baryon asymmetry can be obtained for a wide range of Yukawa
  couplings in eq.~(\ref{yuk})\cite{luty,pluemi,etc}. Further, one may
  ask whether the right order of magnitude of the asymmetry results
  naturally in the leptogenesis scenario. To address this question one
  has to discuss patterns of neutrino mass matrices which determine
  the generated asymmetry.

\section{Neutrino masses and mixings}

  In sect.~1 we argued that a dynamical generation of the cosmological
  baryon asymmetry requires lepton number violation. This is most
  easily realized by adding right-handed neutrinos to the standard
  model. In the context of unified theories one is then led to go
  beyond the SU(5) GUT and to consider SO(10) as smallest unified
  gauge group allowing right-handed neutrinos. In the following we
  shall therefore assume a similar pattern of mixings and mass ratios
  for leptons and quarks\cite{bp}, which is natural in SO(10)
  unification.
  
  Such an ansatz is most transparent in a basis where all mass
  matrices are maximally diagonal. In addition to real mass
  eigenvalues two mixing matrices then appear. One can always choose a
  basis for the lepton fields such that the mass matrices $m_l$ for
  the charged leptons and $M$ for the heavy Majorana neutrinos $N_i$
  are diagonal with real and positive eigenvalues,
  \beq
    m_l=\left(\begin{array}{ccc}m_e&0&0\\0&m_{\m}&0\\0&0&m_{\t}
              \end{array}\right)\qquad
    M=\left(\begin{array}{ccc}M_1&0&0\\0&M_2&0\\0&0&M_3
    \end{array}\right)\;.
  \eeq
  In this basis $m_D$ is a general complex matrix, which can be
  diagonalized by a biunitary transformation. Therefore, we can
  write $m_D$ in the form
  \beq
    m_D=V\,\left(\begin{array}{ccc}
    m_1&0&0\\0&m_2&0\\0&0&m_3\end{array}\right)\,U^{\dag}\;,
  \eeq
  where $V$ and $U$ are unitary matrices and the $m_i$ are real and
  positive. In the absence of a Majorana mass term $V$ and $U$ would 
  correspond to Kobayashi-Maskawa type mixing matrices of left- and 
  right-handed charged currents, respectively.
  
  According to eq.~(\ref{cpasymm}) the $CP$ asymmetry is determined by
  the mixings and phases present in the product $m_D^{\dg}m_D$, where
  the matrix $V$ drops out.  Hence, to leading order, the mixings and
  phases which are responsible for baryogenesis are entirely
  determined by the matrix $U$.  Correspondingly, the mixing matrix
  $K$ in the leptonic charged current, which determines $CP$ violation
  and mixings of the light leptons, depends on mass ratios and mixing
  angles and phases of $U$ and $V$.  This implies that there exists no
  direct connection between the $CP$ violation and generation mixing
  relevant at high and low energies.

  Consider now the mixing matrix $U$. One can factor out five phases,
  \beq
    U=\mbox{e}^{i\g}\,\mbox{e}^{i\l_3\a}\,\mbox{e}^{i\l_8\b}\,U_1\,
    \mbox{e}^{i\l_3\s}\,\mbox{e}^{i\l_8\t}\;,
  \eeq
  where the $\l_i$ are the Gell-Mann matrices. The remaining matrix
  $U_1$ depends on three mixing angles and one phase, like the 
  Kobayashi-Maskawa matrix for quarks. In analogy to the quark mixing
  matrix we choose the Wolfenstein parametrization\cite{wolfenstein}
  as ansatz for $U_1$,
  \beq\label{mm}
    U_1=\left(\begin{array}{ccc}
    1-{\l^2\over2}  &      \l        & A\l^3(\r-i\h)\\[1ex]
        -\l         & 1-{\l^2\over2} & A\l^2 \\[1ex]
    A\l^3(1-\r-i\h) &    -A\l^2      &  1
    \end{array}\right)\;,
  \eeq
  where $A$ and $|\r+i\h|$ are $\co(1)$, while the mixing
  parameter $\l$ is assumed to be small. For the masses $m_i$ and
  $M_i$ we assume the same hierarchy which is observed for up-type quarks,
  \beqa
    m_1=b\l^4m_3\;,&\quad m_2=c\l^2m_3\;,&\quad b,c=\co(1)\\[1ex]
    M_1=B\l^4M_3\;,&\quad M_2=C\l^2M_3\;,&\quad B,C=\co(1)\;.\label{Mmass}
  \eeqa
  For the eigenvalues $m_i$ of the Dirac mass matrix this choice is
  suggested by SO(10) unification. For the masses $M_i$ this is an
  assumption motivated by simplicity. The masses $M_i$ cannot be
  degenerate, because in this case there exists a basis for $\n_R$
  such that $U = 1$, which implies that no baryon asymmetry is
  generated. We shall see in the next section that the precise form of the 
  assumed hierarchy has no influence on the viability of 
  the leptogenesis mechanism. 
  
  The light neutrino masses are given by the seesaw formula
  (\ref{seesaw}). The matrix $K$, which diagonalizes the neutrino mass
  matrix, can be evaluated in powers of $\l$. A straightforward
  calculation gives the following masses for the light neutrino mass
  eigenstates
  \beqa
     m_{\n_e}&=&{b^2\over\left|C+\mbox{e}^{4i\a}\;B\right|}\;\l^4
             \;m_{\n_{\t}}+\co\left(\l^6\right)\label{mne}\\[1ex]
     m_{\n_{\m}}&=&{c^2\left|C+\mbox{e}^{4i\a}\;B\right|\over BC}
             \;\l^2\;m_{\n_{\t}}+\co\left(\l^4\right)\label{mnm}\\[1ex]
     m_{\n_{\t}}&=&{m_3^2\over M_3}+\co\left(\l^4\right)\;.\label{mnt}
  \eeqa
 
  The $CP$-asymmetry in the decay of the lightest right-handed
  neutrino $N_1$ is easily obtained from eqs.~(\ref{cpasymm}) and
  (\ref{mm})-(\ref{Mmass}),  
  \beq
    \ve_1=-\;{1\over16\p}\;{B\;A^2\over c^2+A^2\;|\r+i\h|^2}\;\l^4\;
    {m_3^2\over v^2}\;\mbox{Im}\left[(\r-i\h)^2
    \mbox{e}^{i2(\a+\sqrt{3}\b)}\right]
    \;+\;\co\left(\l^6\right)\;.
  \eeq
  This yields for the magnitude of the $CP$ asymmetry,
  \beq\label{cpa}
    |\ve_1| \leq {1\over16\p}\;{B\;A^2\;|\r+i\h|^2\over c^2+A^2\;|\r+i\h|^2}\;
    \l^4\;{m_3^2\over v^2}\;+\;\co\left(\l^6\right)\;.
  \eeq  
  How close the value of $|\ve_1|$ is to this upper bound depends on
  the phases $\a$, $\b$ and $\arg{(\r+i\h)}$.  For $\l \sim 0.1$ one
  has $\ve_1\sim 10^{-6}\cdot m_3^2/v^2$. Hence, a large value of the
  Yukawa coupling $m_3/v$ will be required by this mechanism of
  baryogenesis. This holds irrespective of the neutrino mixings and the
  heavy neutrino masses.

\newpage
\section{Numerical Results}

\begin{figure}[t]
     \mbox{ }\hspace{-0.7cm}
     \begin{minipage}[t]{8.2cm}
     \mbox{ }\hfill\hspace{1cm}(a)\hfill\mbox{ }
     \epsfig{file=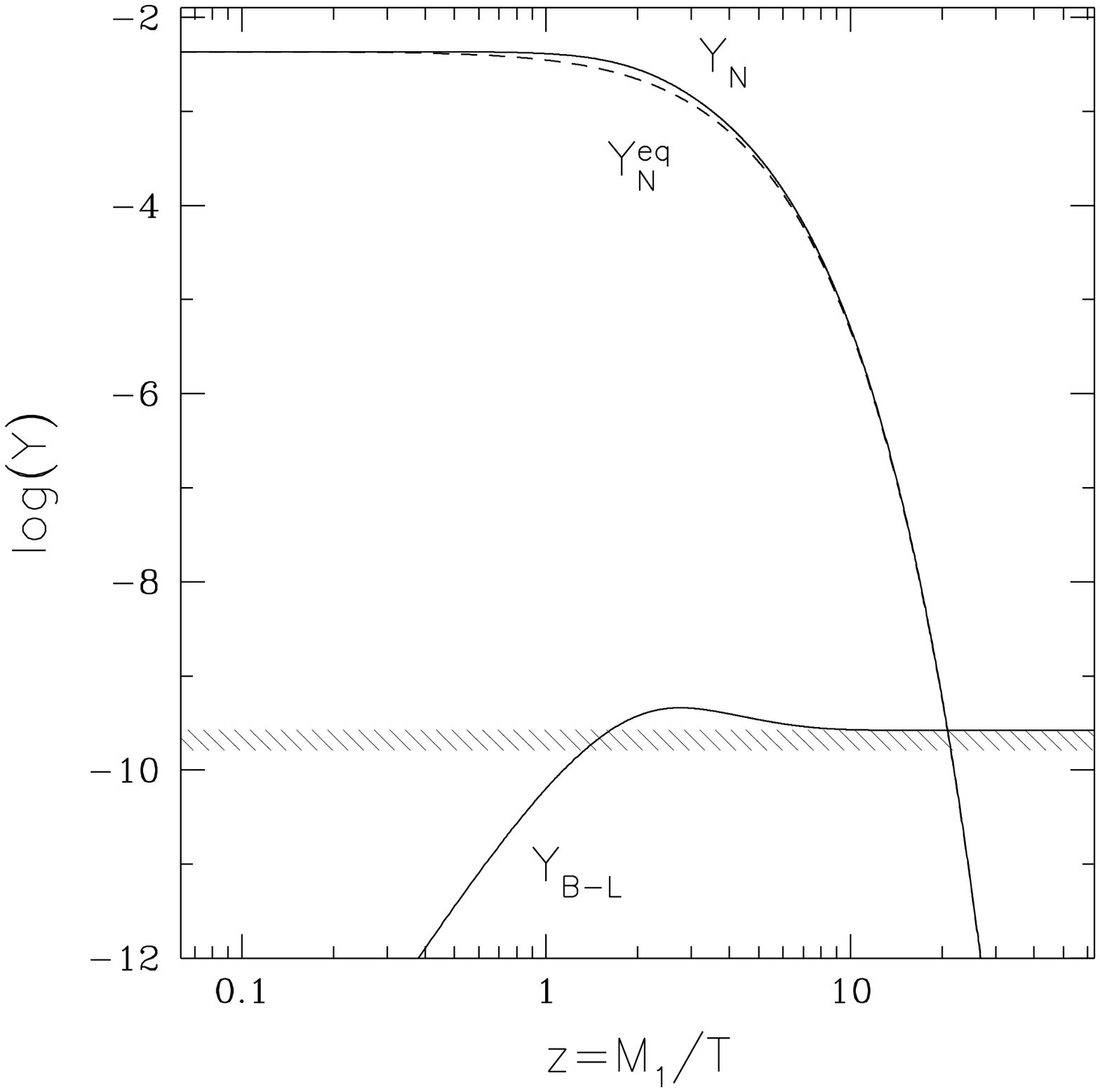,width=8.2cm}
     \end{minipage}
     \hspace{-0.4cm}
     \begin{minipage}[t]{8.2cm}
     \mbox{ }\hfill\hspace{1cm}(b)\hfill\mbox{ }
     \epsfig{file=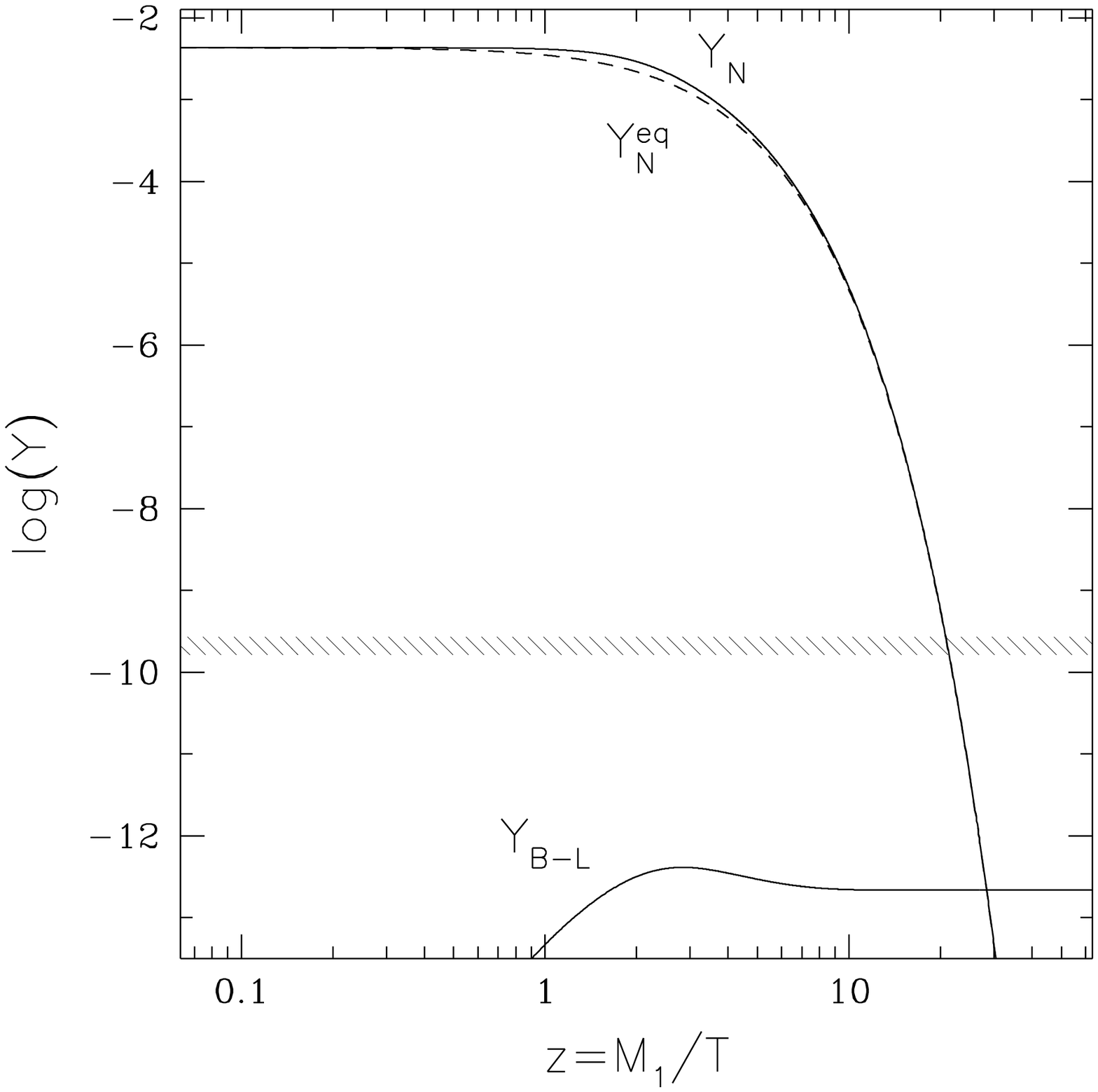,width=8.2cm}
     \end{minipage}  
     \caption{\it Time evolution of the neutrino number density and
     the $B-L$ asymmetry for $\l=0.1$ and for $m_3=m_t$ (a)
     or $m_3=m_b$ (b). The equilibrium distribution for
     $N_1$ is represented by a dashed line, while the hatched area
     shows the measured value for the asymmetry.
     \label{non_susyplot}}
  \end{figure}

  To obtain a numerical value for the produced baryon asymmetry, one
  has to specify the free parameters in the ansatz
  (\ref{mm})-(\ref{Mmass}).  In the following we will use as a
  constraint the value for the $\n_{\m}$-mass which is preferred by
  the MSW explanation\cite{msw} of the solar neutrino 
  deficit\cite{tot},
  \beq
    m_{\n_{\m}}\simeq 3\cdot10^{-3}\;\mbox{eV}\;. \label{msw}
  \eeq
  A generic choice for the free parameters is to take all $\co(1)$
  parameters equal to one and to fix $\l$ to a value which is
  of the same order as the $\l$ parameter of the quark mixing matrix,
  \beq
    A=B=C=b=c=|\r+i\h|\simeq 1\; ,\qquad \l\simeq 0.1\;. \label{p1}
  \eeq
  {}From eqs.~(\ref{mne})-(\ref{mnt}), (\ref{msw}) and (\ref{p1}) one
  now obtains,
  \beq\label{nmasses}
    m_{\n_e}\simeq 8\cdot10^{-6}\;\mbox{eV}\; , \quad
    m_{\n_{\t}}\simeq 0.15\;\mbox{eV}\; .\label{m1}
  \eeq
  Finally, a second mass scale has to be specified. In unified
  theories based on SO$(10)$ the Dirac neutrino mass $m_3$ is
  naturally equal to the top-quark mass,
  \beq\label{3t}
     m_3=m_t\simeq 174\;\mbox{GeV}\;.\label{mtop}
  \eeq
  This determines the masses of the heavy Majorana neutrinos $N_i$,
  \beq
     M_3 \simeq 2\cdot10^{14}\;\mbox{GeV}\; ,\label{M3}
  \eeq  
  and, consequently, $M_1\simeq 2\cdot10^{10}\;\mbox{GeV}$ and
  $M_2\simeq 2\cdot10^{12}\;\mbox{GeV}$. {}From eq.~(\ref{cpa}) one
  obtains the $CP$ asymmetry $|\ve_1| \simeq 10^{-6}$, where we have
  assumed maximal phases. The solution of the Boltzmann equations now
  yields the baryon asymmetry (see fig.~\ref{non_susyplot}a),
  \beq
     Y_B \simeq 9\cdot10^{-11}\; , \label{nonsusy_res1}
  \eeq
  which is indeed the correct order of magnitude! The precise value
  depends on unknown phases.
  
  The large mass $M_3$ of the heavy Majorana neutrino $N_3$
  (cf.~(\ref{M3})), suggests that $B-L$ is already broken at the
  unification scale $\Lambda_{\mbox{\scriptsize GUT}} \sim 10^{16}$
  GeV, without any intermediate scale of symmetry breaking. This large
  value of $M_3$ is a consequence of the choice $m_3 \simeq m_t$. To
  test the sensitivity of the result for $Y_{B-L}$ on this assumption,
  consider as an alternative the choice $m_3 = m_b \simeq 4.5$ GeV, with
  all other parameters remaining unchanged. In this case one has
  $M_3=10^{11}$ GeV and $|\ve_1| = 5\cdot10^{-10}$ for the mass of
  $N_3$ and the $CP$ asymmetry, respectively. Since the maximal $B-L$
  asymmetry is $-\ve_1/g*$, it is clear that the generated asymmetry
  will be too small. The solutions of the Boltzmann equations are
  shown in fig.~\ref{non_susyplot}b. The generated asymmetry, 
  \beq
    Y_B\simeq 8\cdot10^{-14}\;,\label{nonsusy_res2}
  \eeq
  is too small by more than two orders of magnitude. We conclude that
  high values for both masses $m_3$ and $M_3$ are preferred, which is
  natural in SO(10) unification.

\begin{figure}[t]
    \mbox{ }\hfill
    \epsfig{file=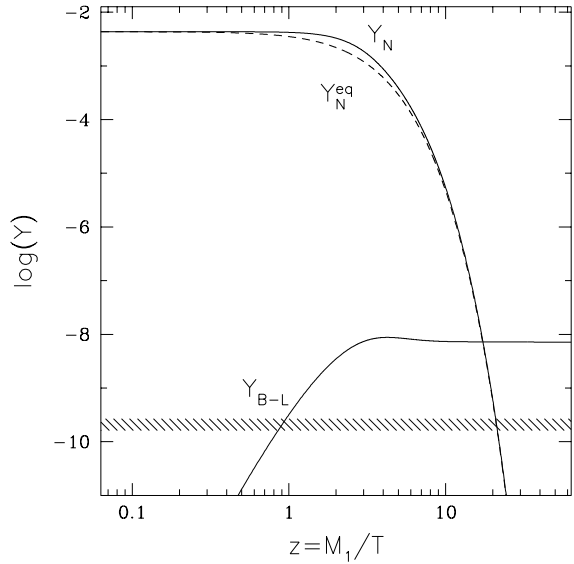,width=8.2cm}
    \hfill\mbox{ }
    \caption{\it Generated lepton asymmetry if one assumes a
        similar mass hierarchy for the right-handed neutrinos and the
        down-type quarks.
        \label{hierarch}}
  \end{figure}

  In eq.~(\ref{Mmass}) we had assumed a mass hierarchy for the heavy
  Majorana neutrinos like for the up-type quarks. One may also
  consider a weaker hierarchy, like for the down-type quarks. This
  corresponds to the choice $B=10$, $C=3$.  Keeping all other
  parameters in eq.~(\ref{p1}) one obtains for the $\n_e$ and
  $\n_{\t}$ masses,
  \beq
    m_{\n_e}\simeq5\cdot10^{-6}\;\mbox{eV}\;,\qquad\qquad
    m_{\n_{\t}}\simeq0.7\;\mbox{eV}\;.\label{nmasses2}
  \eeq
  The large Dirac mass (\ref{3t}) again leads to a large Majorana mass 
  \beq
    M_3 \simeq 4\cdot10^{13}\;\mbox{GeV}\;,
  \eeq  
  and, consequently, $M_1 \simeq 4\cdot10^{10}$ GeV, $M_2 \simeq
  10^{12}$ GeV. {}From eq.~(\ref{cpa}) one obtains the $CP$ asymmetry
  $\ve_1\simeq-10^{-6}$. The corresponding solutions of the Boltzmann
  equations are shown in fig.~\ref{hierarch}. The final baryon
  asymmetry,
  \beq
    Y_B \simeq 2\cdot10^{-9}\;,\label{nonsusy_res3}
  \eeq
  is larger than required, but this value can always be
  lowered by adjusting the unknown phases. Hence, the possibility to
  generate a lepton asymmetry does not depend on the special form of
  the mass hierarchy assumed for the right-handed neutrinos, as long
  as some kind of mass hierarchy exists. 
  
  Models for dark matter involving massive neutrinos favour a
  $\t$-neutrino mass $m_{\n_{\t}} \simeq 5\; \mbox{eV}$\cite{raf},
  which is significantly larger than the value given in (\ref{m1}).
  Such a large value for the $\t$-neutrino mass can be accommodated
  within the ansatz described in this section. However, it does not
  correspond to the simplest choice of parameters and requires some
  fine-tuning. For the mass of the heaviest Majorana neutrino one
  obtains in this case $M_3 \simeq 6\cdot 10^{12}$ GeV.
  
  The recently reported atmospheric neutrino anomaly\cite{tot} may be
  due to neutrino oscillations. The required mass difference and
  mixing angle are $\Delta m^2 \sim 0.005$ eV$^2$ and $\sin^2{2\Theta}
  \sim 1$. The preferred solution for baryogenesis discussed above
  yields (cf.~eq.~(\ref{nmasses})) $m_{\n_\t}^2-m_{\n_\m}^2 \simeq
  0.02$ eV$^2$ which, within the theoretical and experimental
  uncertainties, is certainly consistent with the mass difference
  required by the neutrino oscillation hypothesis. The $\n_\t$-$\n_\m$
  mixing angle is not constrained by leptogenesis and therefore a free
  parameter in principle. The large value needed, however, is against
  the spirit of small generation mixings manifest in the Wolfenstein
  ansatz and would require some special justification.

\section{Supersymmetric extension}

  Without an intermediate scale of symmetry breaking, the unification
  of gauge couplings appears to require low-energy supersymmetry.
  Supersymmetric leptogenesis has already been considered\cite{camp}
  in an approximation where lepton number violating scatterings are
  neglected which inhibit the generation of lepton number. However, a
  full analysis of the mechanism including all the relevant scattering
  processes is necessary in order to get a reliable relation between
  the input parameters and the final asymmetry\cite{pluemi2}. It turns
  out that the lepton number violating scatterings are qualitatively
  more important than in the non-supersymmetric scenario and that they
  can even account for the generation of an equilibrium distribution
  of heavy neutrinos at high temperatures.
  \begin{figure}[t]
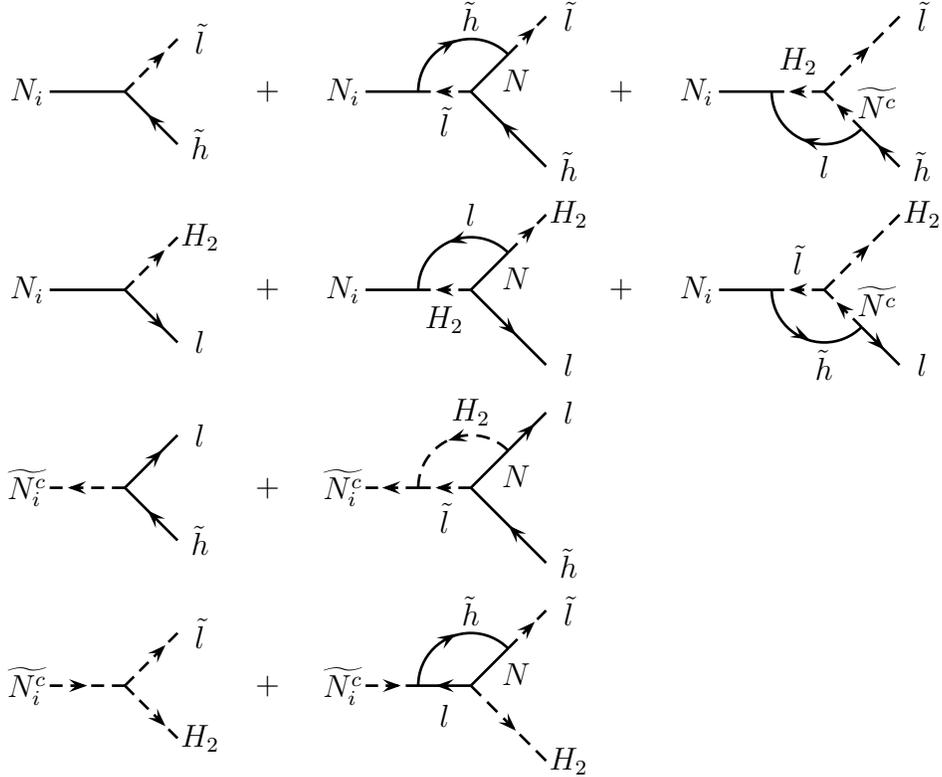

    \begin{center}
\parbox[c]{12.5cm}{
\pspicture(0,0)(3.7,2.6)
\psline[linewidth=1pt](0.6,1.3)(1.6,1.3)
\psline[linewidth=1pt](1.6,1.3)(2.3,0.6)
\psline[linewidth=1pt,linestyle=dashed](1.6,1.3)(2.3,2.0)
\psline[linewidth=1pt]{<-}(1.9,1.0)(2.0,0.9)
\psline[linewidth=1pt]{->}(2.03,1.73)(2.13,1.83)
\rput[cc]{0}(0.3,1.3){$N_i$}
\rput[cc]{0}(2.6,0.6){$\tilde{h}$}
\rput[cc]{0}(2.6,2.0){$\tilde{l}$}
\rput[cc]{0}(3.5,1.3){$+$}
\endpspicture
\pspicture(-0.5,0)(4.2,2.6)
\psline[linewidth=1pt](0.6,1.3)(1.3,1.3)
\psline[linewidth=1pt,linestyle=dashed](1.3,1.3)(2.0,1.3)
\psline[linewidth=1pt](2,1.3)(2.5,1.8)
\psline[linewidth=1pt,linestyle=dashed](2.5,1.8)(3,2.3)
\psline[linewidth=1pt](2,1.3)(3,0.3)
\psarc[linewidth=1pt](2,1.3){0.7}{45}{180}
\psline[linewidth=1pt]{<-}(1.53,1.3)(1.63,1.3)
\psline[linewidth=1pt]{->}(1.7,1.93)(1.8,1.99)
\psline[linewidth=1pt]{->}(2.75,2.05)(2.85,2.15)
\psline[linewidth=1pt]{<-}(2.4,0.9)(2.5,0.8)
\rput[cc]{0}(0.3,1.3){$N_i$}
\rput[cc]{0}(1.65,0.9){$\tilde{l}$}
\rput[cc]{0}(2,2.3){$\tilde{h}$}
\rput[cc]{0}(2.6,1.45){$N$}
\rput[cc]{0}(3.3,2.3){$\tilde{l}$}
\rput[cc]{0}(3.3,0.3){$\tilde{h}$}
\rput[cc]{0}(4.0,1.3){$+$}
\endpspicture
\pspicture(-0.5,0)(3.5,2.6)
\psline[linewidth=1pt](0.6,1.3)(1.3,1.3)
\psline[linewidth=1pt,linestyle=dashed](1.3,1.3)(2.0,1.3)
\psline[linewidth=1pt,linestyle=dashed](2,1.3)(2.5,0.8)
\psline[linewidth=1pt](2.5,0.8)(3,0.3)
\psline[linewidth=1pt,linestyle=dashed](2,1.3)(3,2.3)
\psarc[linewidth=1pt](2,1.3){0.7}{-180}{-45}
\psline[linewidth=1pt]{<-}(1.53,1.3)(1.63,1.3)
\psline[linewidth=1pt]{<-}(1.7,0.67)(1.8,0.64)
\psline[linewidth=1pt]{<-}(2.16,1.14)(2.26,1.04)
\psline[linewidth=1pt]{<-}(2.7,0.6)(2.8,0.5)
\psline[linewidth=1pt]{->}(2.5,1.8)(2.6,1.9)
\rput[cc]{0}(0.3,1.3){$N_i$}
\rput[cc]{0}(1.65,1.7){$H_2$}
\rput[cc]{0}(2,0.3){$l$}
\rput[cc]{0}(2.7,1.15){$\widetilde{N^c}$}
\rput[cc]{0}(3.3,0.3){$\tilde{h}$}
\rput[cc]{0}(3.3,2.3){$\tilde{l}$}
\endpspicture\\
\pspicture(0,0)(3.7,2.6)
\psline[linewidth=1pt](0.6,1.3)(1.6,1.3)
\psline[linewidth=1pt](1.6,1.3)(2.3,0.6)
\psline[linewidth=1pt,linestyle=dashed](1.6,1.3)(2.3,2.0)
\psline[linewidth=1pt]{->}(2.03,0.87)(2.13,0.77)
\psline[linewidth=1pt]{->}(2.03,1.73)(2.13,1.83)
\rput[cc]{0}(0.3,1.3){$N_i$}
\rput[cc]{0}(2.6,0.6){$l$}
\rput[cc]{0}(2.6,2.0){$H_2$}
\rput[cc]{0}(3.5,1.3){$+$}
\endpspicture
\pspicture(-0.5,0)(4.2,2.6)
\psline[linewidth=1pt](0.6,1.3)(1.3,1.3)
\psline[linewidth=1pt,linestyle=dashed](1.3,1.3)(2.0,1.3)
\psline[linewidth=1pt](2,1.3)(2.5,1.8)
\psline[linewidth=1pt,linestyle=dashed](2.5,1.8)(3,2.3)
\psline[linewidth=1pt](2,1.3)(3,0.3)
\psarc[linewidth=1pt](2,1.3){0.7}{45}{180}
\psline[linewidth=1pt]{<-}(1.53,1.3)(1.63,1.3)
\psline[linewidth=1pt]{<-}(1.7,1.93)(1.8,1.96)
\psline[linewidth=1pt]{->}(2.75,2.05)(2.85,2.15)
\psline[linewidth=1pt]{->}(2.5,0.8)(2.6,0.7)
\rput[cc]{0}(0.3,1.3){$N_i$}
\rput[cc]{0}(1.65,0.9){$H_2$}
\rput[cc]{0}(2,2.3){$l$}
\rput[cc]{0}(2.6,1.45){$N$}
\rput[cc]{0}(3.3,2.3){$H_2$}
\rput[cc]{0}(3.3,0.3){$l$}
\rput[cc]{0}(4.0,1.3){$+$}
\endpspicture
\pspicture(-0.5,0)(3.5,2.6)
\psline[linewidth=1pt](0.6,1.3)(1.3,1.3)
\psline[linewidth=1pt,linestyle=dashed](1.3,1.3)(2.0,1.3)
\psline[linewidth=1pt,linestyle=dashed](2,1.3)(2.5,0.8)
\psline[linewidth=1pt](2.5,0.8)(3,0.3)
\psline[linewidth=1pt,linestyle=dashed](2,1.3)(3,2.3)
\psarc[linewidth=1pt](2,1.3){0.7}{-180}{-45}
\psline[linewidth=1pt]{<-}(1.53,1.3)(1.63,1.3)
\psline[linewidth=1pt]{->}(1.7,0.69)(1.8,0.64)
\psline[linewidth=1pt]{<-}(2.16,1.14)(2.26,1.04)
\psline[linewidth=1pt]{->}(2.7,0.6)(2.8,0.5)
\psline[linewidth=1pt]{->}(2.5,1.8)(2.6,1.9)
\rput[cc]{0}(0.3,1.3){$N_i$}
\rput[cc]{0}(1.65,1.7){$\tilde{l}$}
\rput[cc]{0}(2,0.3){$\tilde{h}$}
\rput[cc]{0}(2.7,1.15){$\widetilde{N^c}$}
\rput[cc]{0}(3.3,0.3){$l$}
\rput[cc]{0}(3.3,2.3){$H_2$}
\endpspicture\\
\pspicture(0,0)(3.7,2.6)
\psline[linewidth=1pt,linestyle=dashed](0.6,1.3)(1.6,1.3)
\psline[linewidth=1pt](1.6,1.3)(2.3,2.0)
\psline[linewidth=1pt](1.6,1.3)(2.3,0.6)
\psline[linewidth=1pt]{<-}(0.84,1.3)(0.94,1.3)
\psline[linewidth=1pt]{<-}(1.9,1.0)(2.0,0.9)
\psline[linewidth=1pt]{->}(2.0,1.7)(2.1,1.8)
\rput[cc]{0}(0.3,1.3){$\sni$}
\rput[cc]{0}(2.6,0.6){$\tilde{h}$}
\rput[cc]{0}(2.6,2.0){$l$}
\rput[cc]{0}(3.5,1.3){$+$}
\endpspicture
\pspicture(-0.5,0)(3.5,2.6)
\psline[linewidth=1pt,linestyle=dashed](0.6,1.3)(1.3,1.3)
\psline[linewidth=1pt,linestyle=dashed](1.3,1.3)(2.0,1.3)
\psline[linewidth=1pt](2,1.3)(2.5,1.8)
\psline[linewidth=1pt](2.5,1.8)(3,2.3)
\psline[linewidth=1pt](2,1.3)(3,0.3)
\psarc[linewidth=1pt,linestyle=dashed](2,1.3){0.7}{45}{180}
\psline[linewidth=1pt]{<-}(0.82,1.3)(0.92,1.3)
\psline[linewidth=1pt]{<-}(1.53,1.3)(1.63,1.3)
\psline[linewidth=1pt]{<-}(1.7,1.93)(1.8,1.96)
\psline[linewidth=1pt]{->}(2.75,2.05)(2.85,2.15)
\psline[linewidth=1pt]{<-}(2.5,0.8)(2.6,0.7)
\rput[cc]{0}(0.3,1.3){$\sni$}
\rput[cc]{0}(1.65,0.9){$\tilde{l}$}
\rput[cc]{0}(2,2.3){$H_2$}
\rput[cc]{0}(2.6,1.45){$N$}
\rput[cc]{0}(3.3,2.3){$l$}
\rput[cc]{0}(3.3,0.3){$\tilde{h}$}
\endpspicture\\
\pspicture(0,0)(3.7,2.6)
\psline[linewidth=1pt,linestyle=dashed](0.6,1.3)(1.6,1.3)
\psline[linewidth=1pt,linestyle=dashed](1.6,1.3)(2.3,2.0)
\psline[linewidth=1pt,linestyle=dashed](1.6,1.3)(2.3,0.6)
\psline[linewidth=1pt]{->}(1.0,1.3)(1.1,1.3)
\psline[linewidth=1pt]{->}(2.04,1.74)(2.14,1.84)
\psline[linewidth=1pt]{->}(2.04,0.86)(2.14,0.76)
\rput[cc]{0}(0.3,1.3){$\sni$}
\rput[cc]{0}(2.6,2.0){$\tilde{l}$}
\rput[cc]{0}(2.6,0.6){$H_2$}
\rput[cc]{0}(3.5,1.3){$+$}
\endpspicture
\pspicture(-0.5,0)(3.5,2.6)
\psline[linewidth=1pt,linestyle=dashed](0.6,1.3)(1.3,1.3)
\psline[linewidth=1pt](1.3,1.3)(2.0,1.3)
\psline[linewidth=1pt](2,1.3)(2.5,1.8)
\psline[linewidth=1pt,linestyle=dashed](2.5,1.8)(3,2.3)
\psline[linewidth=1pt,linestyle=dashed](2,1.3)(3,0.3)
\psarc[linewidth=1pt](2,1.3){0.7}{45}{180}
\psline[linewidth=1pt]{->}(0.97,1.3)(1.07,1.3)
\psline[linewidth=1pt]{<-}(1.53,1.3)(1.63,1.3)
\psline[linewidth=1pt]{->}(1.7,1.93)(1.8,1.99)
\psline[linewidth=1pt]{->}(2.75,2.05)(2.85,2.15)
\psline[linewidth=1pt]{->}(2.5,0.8)(2.6,0.7)
\rput[cc]{0}(0.3,1.3){$\sni$}
\rput[cc]{0}(1.65,0.9){$l$}
\rput[cc]{0}(2,2.3){$\tilde{h}$}
\rput[cc]{0}(2.6,1.45){$N$}
\rput[cc]{0}(3.3,2.3){$\tilde{l}$}
\rput[cc]{0}(3.3,0.3){$H_2$}
\endpspicture
}
\end{center}
    \caption{\it Decay modes of the right-handed Majorana neutrinos
                 and their scalar partners in the supersymmetric
                 scenario. \label{susy_decay}}
  \end{figure}
  
  The supersymmetric generalization of the lagrangian (\ref{yuk}) is
  the superpotential
  \beq
    W = {1\over2}N^cMN^c + \m H_1\e H_2 + H_1 \e L \l_l E^c 
        + H_2 \e L \l_{\n} N^c\;,
  \eeq
  where, in the usual notation, $H_1$, $H_2$, $L$, $E^c$ and $N^c$ are
  chiral superfields describing spin-0 and spin-${1\over 2}$ fields.
  The basis for the lepton fields can be chosen as in the
  non-supersymmetric case.  The vacuum expectation values
  $v_1=\left\langle H_1\right\rangle$ and $v_2=\left\langle
  H_2\right\rangle$ of the two neutral Higgs fields generate Dirac
  masses for the leptons and their scalar partners,
  \beq
    m_l=\l_l\;v_1\quad,\qquad m_D=\l_{\n}\;v_2\;.
  \eeq

  The heavy neutrinos and their scalar partners can decay into
  various final states (cf.~fig.~\ref{susy_decay}). At tree level,
  the decay widths read,
  \beqa
    \G_{rs}\Big(N_i\to\wt{l}+\wt{h}^c\;\Big)
       =\G_{rs}\Big(N_i\to l+H_2\Big)
       &=&{1\over16\p}\;{\mmii\over v_2^2}\ M_i\;,
         \label{decay1}\\[1ex]
    \G_{rs}\Big(\sni\to\wt{l}+H_2\Big)
       =\G_{rs}\Big(\wt{N_i}\to l+\wt{h}^c\;\Big)
       &=&{1\over8\p}\;{\mmii\over v_2^2}\ M_i\;.
        \label{decay2}
  \eeqa
  The $CP$ asymmetry in each of the decay channels is given
  by\cite{covi} 
  \beqa
     \ve_i&=& -{1\over8\p v_2^2}\;{1\over\mmii}\sum\limits_j\mbox{Im}
        \left[(m_D^{\dg}m_D)^2_{ji}\right]\;
        g\left({M_j^2\over M_i^2}\right)\\[1ex]
      &&\qquad g(x)=\sqrt{x}\;\mbox{ln}\left({1+x\over x}\right)\;.
  \eeqa
  It arises through interference of tree level and one-loop diagrams
  shown in fig.~\ref{susy_decay}.  In the case of a mass hierarchy,
  $M_j\gg M_i$, the $CP$ asymmetry is twice as large as in the
  non-supersymmetric case.
  
  Like in the non-supersymmetric scenario lepton number violating
  scatterings mediated by a heavy (s)neutrino have to be included in a
  consistent analysis, since they can easily reduce the generated
  asymmetry by two orders of magnitude\cite{pluemi2}. A very
  interesting new feature of the supersymmetric model is that the
  (s)neutrino (s)top scatterings are strong enough to bring the
  neutrinos into thermal equilibrium at high temperatures. Hence, an
  equilibrium distribution can be reached for temperatures far below
  the masses of heavy gauge bosons.

  \begin{figure}[t]
     \begin{center}
     \epsfig{file=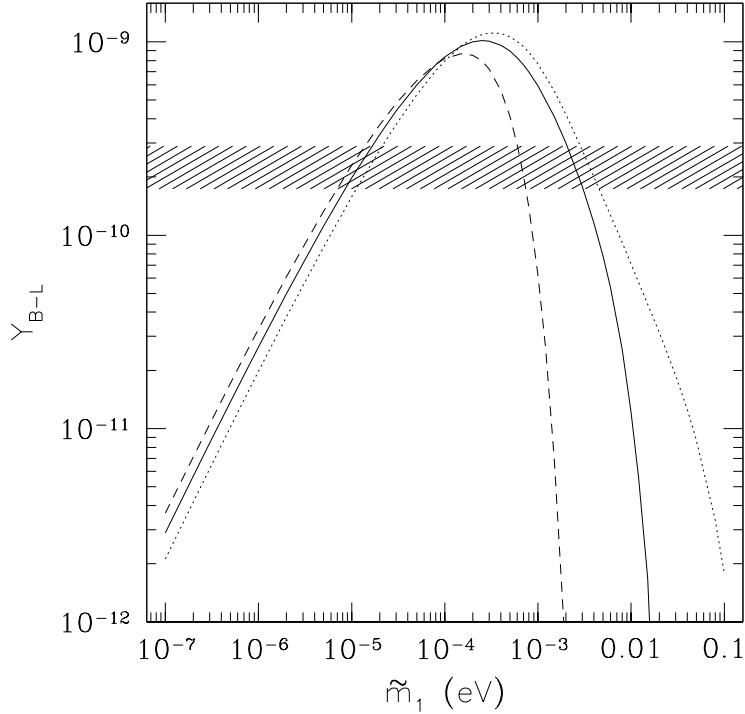,width=10cm}
     \end{center}
     \caption{\it The generated $(B-L)$ asymmetry for
       $M_1=10^8\;$GeV (dotted line), $M_1=10^{10}\;$GeV (solid line)
       and $M_1=10^{12}\;$GeV (dashed line). The hatched area shows
       the measured value for the asymmetry.
       \label{susy_res}}
  \end{figure}
  
  {}From the discussion of the out-of-equilibrium condition in sect.~1
  we know that the generated baryon asymmetry is very sensitive to the
  decay width $\Gamma_1$ of $N_1$, and therefore to $\mmoo$. In fact,
  it turns out that the asymmetry essentially depends on the ratio
  \beq
    \wt{m}_1 = {\mmoo\over M_1}\;. 
  \eeq
  For the mass matrices discussed in sect.~4 $\wt{m}_1$ is of the same
  order as the muon neutrino mass. One easily verifies,
  \beq
    \wt{m}_1 = {C(c^2+A^2|\r+i\eta|^2)\over 
                c^2 \left|C + e^{4i\alpha} B\right|}\,m_{\n_\m}\;
                + \co(\l^2)\;.
    \label{m1t}
  \eeq
  
  In fig.~\ref{susy_res} we have plotted the generated lepton
  asymmetry as function of $\wt{m}_1$ for three different values of
  $M_1$, where we have assumed the hierarchy $M_2^2=10^3\;M_1^2$,
  $M_3^2=10^6\;M_1^2$ and the $CP$ asymmetry $\ve_1=-10^{-6}$.
  
  Fig.~\ref{susy_res} demonstrates, first of all, that in the whole
  parameter range the generated asymmetry is much smaller than the
  value $4\cdot10^{-9}$ which one obtains from the naive estimate
  (\ref{esti}), neglecting lepton number violating scattering
  processes.  For small $\wt{m}_1$ the reason is that the Yukawa
  interactions are too weak to bring the neutrinos into equilibrium at
  high temperatures.  For large $\wt{m}_1$, on the other hand, the
  lepton number violating scatterings wash out a large part of the
  generated asymmetry at temperatures $T<M_1$.
  
  Baryogenesis is possible in the range
  \beq
    10^{-5}\;\mbox{eV}\;\ltap\;\wt{m}_1\;\ltap\;
    5\cdot10^{-3}\;\mbox{eV}\;.
  \eeq
  This result is independent of any assumptions on the mass matrices, 
  in particular it is not a consequence of the ansatz discussed in 
  sect.~5. This ansatz just implies (cf.~(\ref{m1t}))
  \beq
    \wt{m}_1\ \simeq\ m_{\n_\m}\;.
  \eeq
  It is very interesting that the $\n_{\m}$-mass preferred by the 
  MSW explanation of the solar neutrino deficit lies indeed in the
  interval allowed by baryogenesis according to fig.~\ref{susy_res}. 

  \begin{figure}[t]
      \mbox{ }\hspace{-0.7cm}
      \begin{minipage}[t]{8.1cm}
        \begin{center}\hspace{1cm}(a)\end{center}
        \epsfig{file=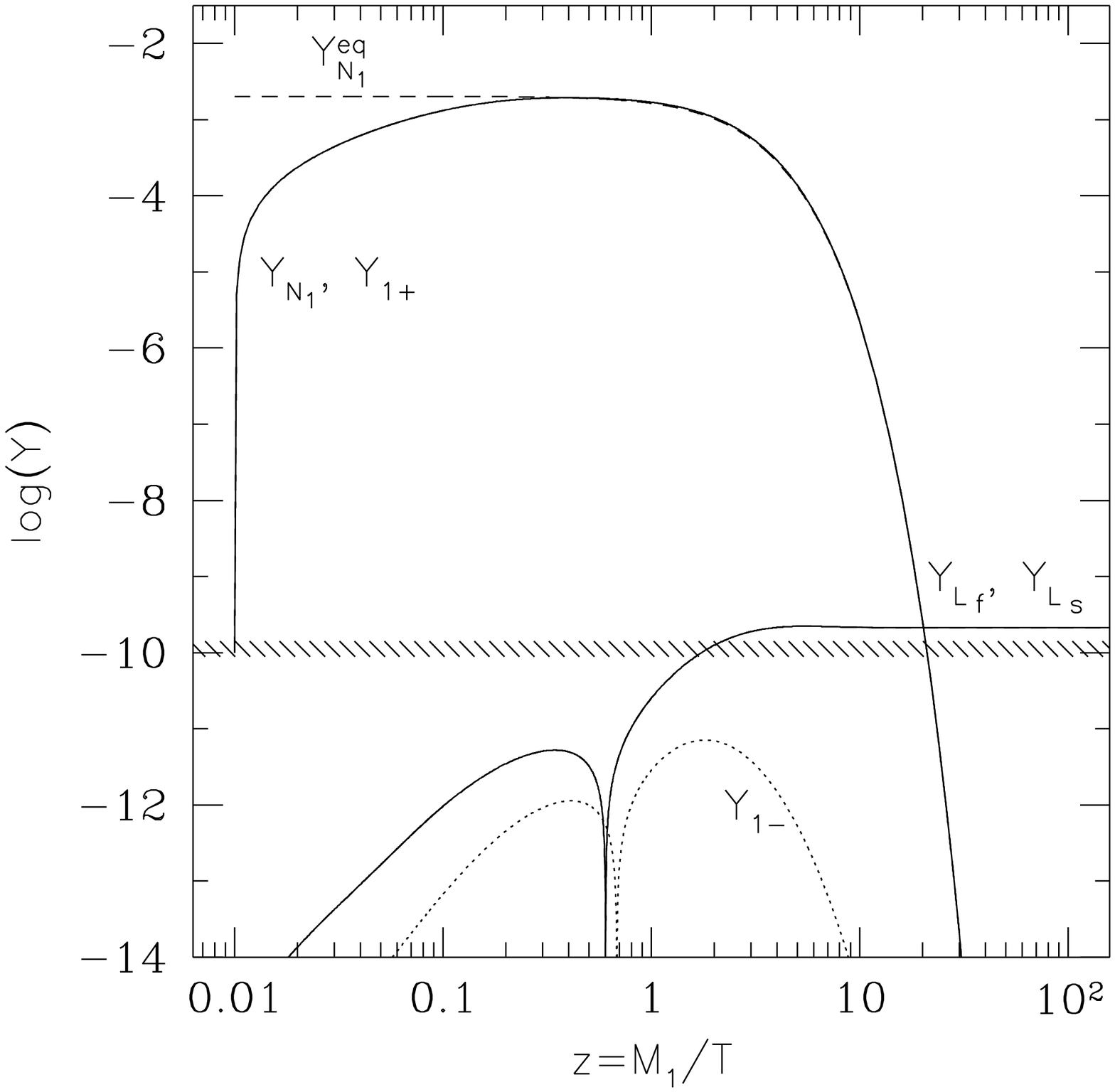,width=8.1cm}
      \end{minipage}
      \hspace{-0.3cm}
      \begin{minipage}[t]{8.1cm}
        \begin{center}\hspace{1cm}(b)\end{center}
        \epsfig{file=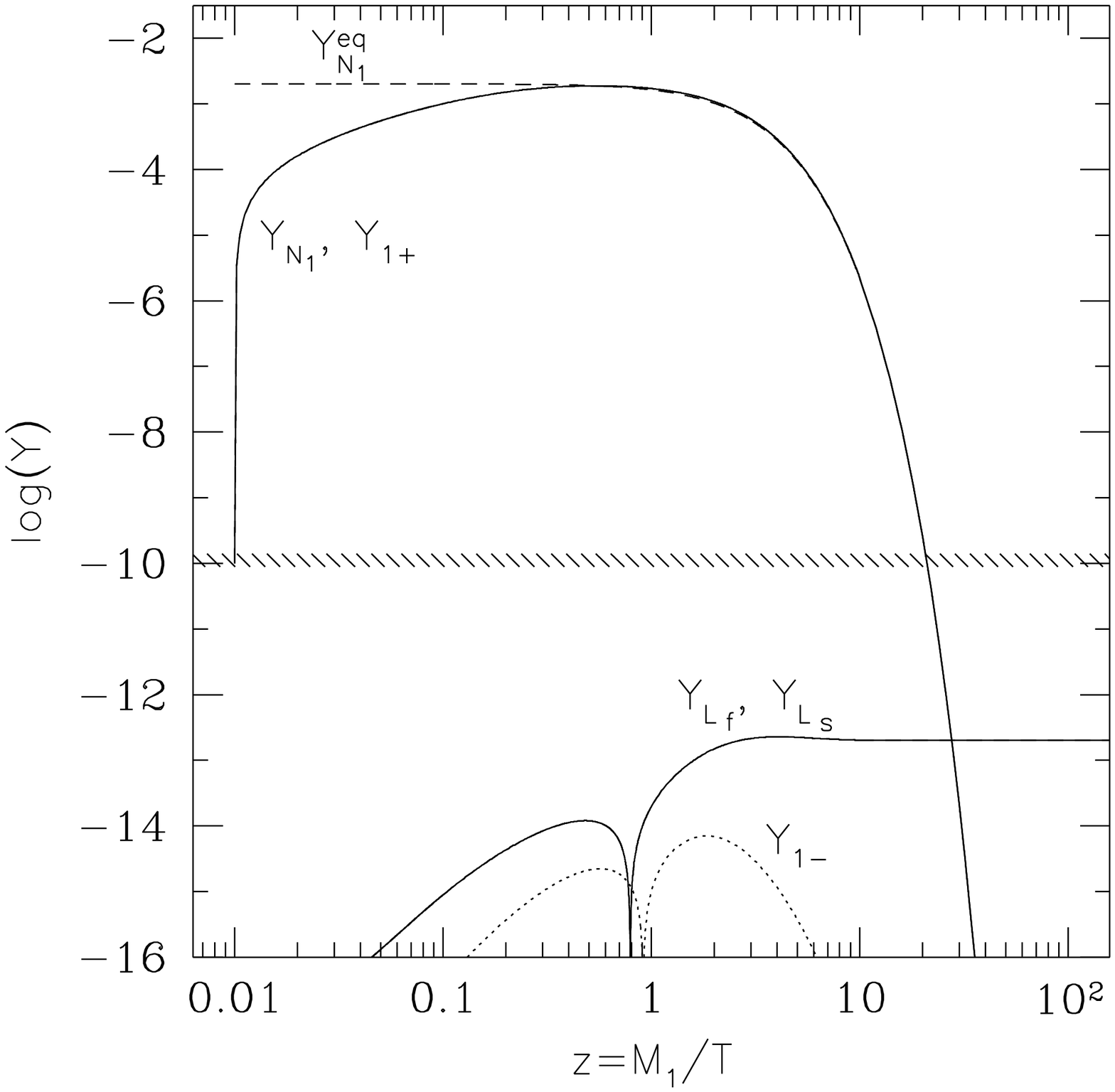,width=8.1cm}
     \end{minipage}  
       \caption{\it Generated asymmetry if one assumes a similar
        pattern of masses and mixings for the leptons and the
        quarks. In both figures we have $\l=0.1$ and $m_3=m_t$ (a) and
        $m_3=m_b$ (b).
       \label{sol2fig}}
  \end{figure}
  
  Consider now again the simplest choice of parameters given by
  eqs.~({\ref{msw})-(\ref{mtop}). The corresponding generated lepton
  asymmetries are shown in fig.~\ref{sol2fig}a. $Y_{L_f}$ and
  $Y_{L_s}$ denote the absolute values of the asymmetries stored in
  leptons and their scalar partners, respectively. They are related
  to the baryon asymmetry by
  \beq
    Y_B = - {8\over 23}\ Y_L\quad, \qquad Y_L = Y_{L_f} + Y_{L_s}\;.
  \eeq 
  $Y_{N_1}$ is the number of heavy neutrinos per comoving volume
  element, and
  \beq
    Y_{1\pm} = Y_{\snone}\pm Y_{\wt{N}_1},
  \eeq
  where $Y_{\snone}$ is the number of scalar neutrinos per comoving
  volume element. As fig.~(\ref{sol2fig}a) shows, the generated baryon
  asymmetry has the correct order of magnitude, like in the
  non-supersymmetric case,
  \beq
    Y_B\simeq10^{-10}\;.\label{susy_res1}
  \eeq  
    
  Lowering the Dirac mass scale of the neutrinos to the bottom-quark
  scale has again dramatic consequences (cf.~fig.~\ref{sol2fig}b). The
  baryon asymmetry is reduced by three orders of magnitude
  \beq
    Y_B\simeq 10^{-13}\;. \label{susy_res2}
  \eeq
  Hence, like in the non-supersymmetric scenario, large values for both
  masses $m_3$ and $M_3$ are necessary.
  
  \begin{figure}[t]
    \mbox{ }\hfill
    \epsfig{file=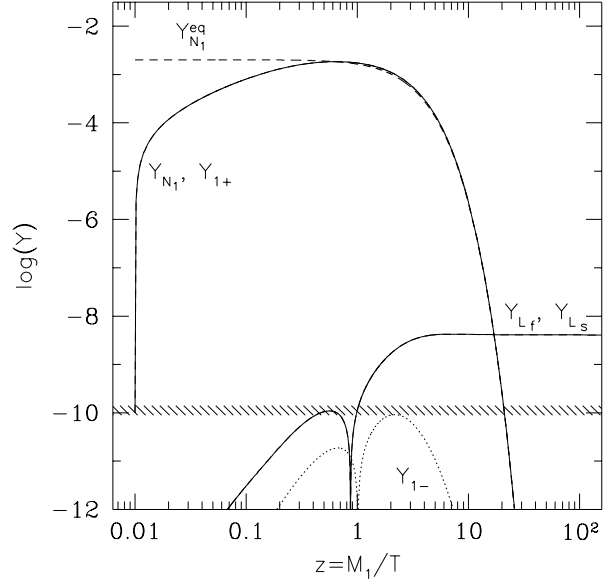,width=8.2cm}
    \hfill\mbox{ }
    \caption{\it Generated lepton asymmetry if one assumes a
        similar mass hierarchy for the right-handed neutrinos and the
        down-type quarks.
        \label{susyhierarch}}
  \end{figure}
  Again like in the non-supersymmetric case the result is insensitive
  to the precise form of the assumed hierarchy for the right-handed
  neutrino masses. Repeating the calculation with the parameter choice
  corresponding to eq.~(\ref{nmasses2}) yields the results shown in
  fig.~\ref{susyhierarch}. The final asymmetry reads
  \beq
    Y_B\simeq 3\cdot10^{-9}\;. \label{susy_res3}
  \eeq

  Comparing the results (\ref{susy_res1}), (\ref{susy_res2}) and
  (\ref{susy_res3}) with their non-supersymmetric counterparts
  (\ref{nonsusy_res1}), (\ref{nonsusy_res2}) and (\ref{nonsusy_res3}),
  one sees that the larger $CP$ asymmetry and the additional
  contributions from the sneutrino decays in the supersymmetric
  scenario are compensated by the wash-out processes which are
  stronger than in the non-supersymmetric case. The final asymmetries
  are the same in the non-supersymmetric and in the supersymmetric
  case.

\section{Summary}

  Anomalous electroweak $B+L$ violating processes are in thermal
  equilibrium in the high-temperature phase of the standard model.  As
  a consequence, asymmetries in baryon and lepton number are related
  at high temperatures, and the cosmological baryon asymmetry can be
  generated from a primordial lepton asymmetry. Necessary ingredients
  are right-handed neutrinos and Majorana masses, which occur
  naturally in SO(10) unification.
  
  The baryon asymmetry can be computed by standard methods based on
  Boltzmann equations. In a consistent analysis lepton number
  violating scatterings have to be taken into account, since they can
  erase a large part of the asymmetry. In supersymmetric scenarios
  these scatterings are sufficient to generate an initial equilibrium
  distribution of heavy Majorana neutrinos.
  
  Baryogenesis implies stringent constraints on the light neutrino
  masses.  Assuming a similar pattern of mixings and masses for
  neutrinos and up-type quarks, as suggested by SO(10) unification,
  the observed asymmetry is obtained without any fine tuning. The
  $\n_{\m}$ mass is predicted in a range consistent with the MSW
  solution of the solar neutrino problem. $B-L$ is broken at the
  unification scale. The baryogenesis scale is given by the mass of
  the lightest of the heavy Majorana neutrinos, which is much lower
  and consistent with constraints from inflation and the gravitino
  abundance.
  
  As our discussion illustrates, the cosmological baryon asymmetry is
  closely related to neutrino properties. Already the existence of a
  baryon asymmetry is a strong argument for lepton number violation
  and Majorana neutrino masses. Together with further information
  about neutrino properties from high-energy physics and astrophysics,
  the theory of the baryon asymmetry will give us new insights into
  physics beyond the standard model.

\mbox{ }\vspace{4ex}\\
\noindent
\setlength{\parskip}{1ex}
{\bf Acknowledgments}\\
\mbox{ }\\    
One of us (W.B.) would like to thank H.~J.~de~Vega and N.~S\'anchez for
organizing an enjoyable and stimulating colloquium.
\mbox{ }\\\noindent

\mbox{ }\vspace{3ex}\\

\end{document}